\documentclass[a4paper,dvips,12pt]{article}
\input{axodraw.sty}
\usepackage{amsmath,amssymb,exscale}   
\usepackage{array,multicol}
\usepackage{afterpage,float,flafter}   
\usepackage{epsfig,rotating,pifont,fancybox}   
\usepackage{cite}
\makeatletter   
\@addtoreset{equation}{section}   
\makeatother   
   
\newcommand{\bea}{\begin{eqnarray}}   
\newcommand{\eea}{\end{eqnarray}}   
   
\newcommand{\NPB}[3]{\emph{ Nucl.~Phys.} \textbf{B#1} (#2) #3}   
\newcommand{\PLB}[3]{\emph{ Phys.~Lett.} \textbf{B#1} (#2) #3}   
\newcommand{\PRD}[3]{\emph{ Phys.~Rev.} \textbf{D#1} (#2) #3}

\newcommand{\IJMPA}[3]{\emph{ Int. J. Mod. Phys. A} \textbf{#1} (#2) #3}  
\newcommand{\Tr}{\mathop{\rm Tr}}   
   
\newcommand{\dalpha}{{\dot\alpha}}

\def\ctu(#1,#2){
\BCirc(#1,#2)4
\put(0,0){
}
\put(0,0){
}
}
\def\ctq(#1,#2){
\BCirc(#1,#2)4
\put(0,0){
}
\put(0,0){
}
}

\title{   
\vspace*{-0.8cm}   
\begin{flushright}   
\normalsize{      
IEM-FT-216/01\\
IFT-UAM/CSIC-01-23\\   
\texttt{hep-ph/0107233}}\\ 
\end{flushright}    
\vspace{1cm}
\Large\textbf{Two-loop Higgs mass in supersymmetric 
Kaluza-Klein theories~\footnote{Work 
supported in part by CICYT, Spain, under contract AEN98-0816,
and by EU under contracts HPRN-CT-2000-00152 and HPRN-CT-2000-00148.}}
\vspace*{.5cm}
\author{\large
{\bf A.~Delgado, G.~v.~Gersdorff  and M.~Quir\'{o}s}\\ \\
\emph{Instituto de Estructura de la Materia (CSIC), Serrano 123,}\\
\emph{E-28006-Madrid, Spain.}}}
\date{}   
\begin{document}
\maketitle
\thispagestyle{empty}
\vspace*{.5cm}

\begin{abstract}\noindent
We perform an explicit two-loop computation of the 
Higgs mass in a five-dimensional supersymmetric
theory compactified on the orbifold $S^1/\mathbb{Z}_2$, with superpotential
localized on a fixed-point brane and supersymmetry broken in 
the bulk of the extra dimension by Scherk-Schwarz boundary conditions.
At one-loop, the Higgs mass
is finite and proportional to $ 1/R^{\,2}$, $R$ being the radius of
the extra dimension. The two-loop corrections contain finite
($\propto 1/R^{\,2}$) and linearly divergent
($\propto\Lambda/R$) terms, where $\Lambda$ is the cutoff of the
five-dimensional theory. The divergent terms are cancelled by diagrams
containing one-loop wave-function renormalization counterterms, i.e.
by linear threshold effects of Yukawa (and gauge) couplings, 
at the scale $1/R$. As a result there is no counterterm for the
Higgs mass, which is finite and calculable to the considered order.
 
\end{abstract}
\vspace{2.cm}   
   
\begin{flushleft}   
July 2001 \\   
\end{flushleft}   
\newpage

\section{Introduction} 
\label{introduction}
There has been recently a lot of effort in order to provide an explanation for
electroweak symmetry breaking (EWSB) by means of supersymmetric
theories with one compact (radius $R$) extra 
dimension~\cite{antoniadis}-\cite{quiros}. Such theories arise
naturally as low energy limits of string theories~\cite{string-reviews} 
and are to be understood as effective theories, when all heavy string
modes are integrated out, with an 
ultraviolet (UV) cutoff $\Lambda$
of the order of the string scale $M_s$. 

A common claim of these models is that electroweak symmetry breaking
(EWSB) is triggered by radiative
corrections which are insensitive to the scale $\Lambda$.  
Some one loop calculations of the effective potential,
or the Higgs mass, have been performed 
in order to establish this result in 
field~\cite{antoniadis}-\cite{quiros}, and 
string~\cite{string-toy} theory, and also some general
arguments have been given~\cite{barbieri3,Nomura01,Weiner01,Masiero01}. 

One uses typically
the so called Kaluza-Klein (KK) regularization which exploits the fact
that sums over KK modes (integration over the fifth momentum) turn out
to be finite (this is already known from finite temperature field
theories).  Since the summation goes over modes with masses far beyond
the cutoff of the effective theory, there have been some doubts on the
validity of such a regularization~\cite{nilles}. However it has been
shown that several cutoffs which respect all symmetries of the theory
lead to a $\Lambda$-insensitive result at one-loop, which equals the one
obtained in the KK regularization~\cite{Delgado01,Contino01}. This is a
very interesting conclusion since it means that EWSB does not depend
on the microscopic theory at all and we have an effective theory with
an unexpected amount of predictability. However, since no rigorous
proof has yet been given which guarantees the independence of EWSB on
the high energy theory, it seems worthwhile to explicitely check this in
a two loop calculation.

In this article we will perform a two-loop calculation 
of the Higgs mass from the top/stop
sector of the model presented in Ref.~\cite{quiros}. It displays the
way how supersymmetric cancellations survive the soft breaking. There
remains a linear sensitivity to the cutoff which is however entirely
absorbed by the supersymmetric running of the Yukawa
coupling~\footnote{In the supersymmetric limit there is only a
wave-function renormalization for the fields which is absorbed by the
running of the Yukawa coupling. This running is of course linear for
$D=5$.}. As a consequence there is no mass counterterm to
the considered order and the Higgs mass is calculable and finite.
These features would remain in the two-loop calculation of the Higgs
mass when the gauge sector is included.

The outline of this paper is as follows. In section 2 we will summarize the
main features of the five-dimensional (5D) supersymmetric model that will
be used in the two-loop calculation. In particular the four dimensional
lagrangian, after integration over the fifth dimension, with all relevant
interactions will be explicitly provided. The corresponding Feynman rules can
be easily deduced from the Lagrangian. In section 3 we review, for 
completeness, the diagrammatic calculation of the one-loop Higgs mass, and the
two-loop contribution is provided in section 4, where the relevant one-loop
counterterms due to the wave-function renormalization of the superfields, 
and their contribution to the two-loop Higgs mass, is computed. Our 
conclusions are drawn in section 5.

\section{The model}

The model we will use for the calculation is the one of Ref.~\cite{quiros}.
The 5D space-time is compactified on the orbifold 
$\mathcal{M}\times S^1/\mathbb{Z}_2$ which has two fixed points at
$y=0,\ell$ ($\ell=\pi R$ is the segment length) where two 3-branes are located.
There are two types of fields: those living in the bulk of the extra 
dimension ($U$-states), organized in $N=2$ supermultiplets, and those living
on the branes and localized at the fixed points ($T$-states), 
organized in $N=1$
supermultiplets. $U$-states are: gauge fields, which appear in $N=2$ vector
multiplets $\mathbb{V}=(V_\mu,\lambda_1;\, \Sigma+iV_5,\lambda_2)$ and 
transform under the adjoint representation of the $SU(3)\times SU(2)_L\times
U(1)_Y$, and $SU(2)_L$ singlets, which appear in $N=2$ hypermultiplets,
$\mathbb{U}=(U,u;U^\prime,u^\prime)$, $\mathbb{D}=(D,d;D^\prime,d^\prime)$
and $\mathbb{E}=(E,e;E^\prime,e^\prime)$. Even and odd $N=1$ superfields under 
$\mathbb{Z}_2$ are separated by a semicolon as $(even;\, odd)$. $T$-states
localized at the $y=0$ brane
are the $SU(2)_L$ doublets, $H_{1,2}=(H_{1,2},h_{1,2})$, $Q=(Q,q)$ and
$L=(L,\ell)$~\footnote{We are using the same notation, upper case symbols, for
the $N=1$ chiral superfields and their scalar components, while lower case
symbols are reserved for the fermionic components, e.g. $U=(U,u)$.}. This
field assignment allows for a superpotential involving the top/stop sector
as
\begin{equation}
\label{superpotencial}
W=h_t^{(5)} Q\, H_2\, U \delta(y)
\end{equation}
which triggers EWSB at one-loop as was shown in Ref.~\cite{quiros}. In the 
rest of this paper we will consider the leading contribution from the top-stop
sector and neglect, in explicit calculations, the contributions from gauge 
couplings. The latter can be easily included and will not modify qualitatively
the result of this paper. 

Let us express the top/stop sector of the theory in terms of
four-dimensional $N=1$ superfields, using the formalism of
Ref.~\cite{Arkani1}. 
\begin{align}
\begin{split}
{\cal L}=&\left( U^\dagger U +U^{\prime\dagger}U'+Q^\dagger Q\delta(y) 
+  H_2^\dagger H_2\delta(y)\right)_{\theta\theta\bar\theta\bar\theta}\\
&+ \left(
U\partial_yU'
+h_t^{(5)}UQH_2\delta(y)+\text{h.c.} \right)_{\theta\theta} 
\label{Superfields}
\end{split}
\end{align}
The superpotential term is given by the second expression and depends
on the chiral superfields $Q,\ H_2,\ U$ and $\partial_yU'$. The
inclusion of the derivative term, on top of the superpotential given in
(\ref{superpotencial}), is pure
convention and it is motivated by the fact that, from a four-dimensional
viewpoint, it constitutes a mass term for the modes. Notice that we have
not included a $\mu H_2 H_1 \delta(y)$ term in the superpotential. The
$\mu$ parameter is a supersymmetric mass and thus its presence is 
completely decoupled from the actual calculation
of the possible soft supersymmetry Higgs
mass counterterms, while it makes the calculation more involved.
For that reason we have decided to work in the (phenomenologically unrealistic)
case $\mu=0$. In section 5 we will make some comments about the impact, in the
present calculation, of the $\mu$-term. We can anticipate, not unexpectedly,
that the final result (in fact the absence of a mass counterterm) will be 
unchanged by the presence of the $\mu$-term.

In terms of component fields we have
\begin{align}
\begin{split}
\label{l5}
\mathcal{L}=
&\partial_\mu U^\dagger\partial^\mu U+
\partial_\mu U^{\prime\dagger}\partial^\mu U'+
\partial_\mu Q^\dagger\partial^\mu Q\delta(y)+
\partial_\mu H_2^\dagger\partial^\mu H_2\delta(y)
\\
&+i\bar u \bar\sigma^\mu\partial_\mu u
+i\bar u' \bar\sigma^\mu\partial_\mu u'
+i\bar q \bar\sigma^\mu\partial_\mu q\delta(y)
+i\bar h_2 \bar\sigma^\mu\partial_\mu h_2\delta(y)
\\
&-|F_	U|^2-|F_{U'}|^2-|F_Q|^2\delta(y)-|F_{H_2}|^2\delta(y)
\\&+
\frac{\partial W}
{\partial U}F_U
+\frac{\partial W}
{\partial U'}F_{U'}
+\frac{\partial W}{\partial Q}F_Q
+\frac{\partial W}{\partial H_2}F_{H_2}
\\&
- \frac{\partial^2 W}
{\partial U \partial U'}uu'
- \frac{\partial^2 W}
{\partial Q \partial U}qu
- \frac{\partial^2 W}
{\partial Q \partial H_2}qh_2
- \frac{\partial^2 W}
{\partial H_2 \partial U}h_2u
+\text{h.c.}
\end{split}
\end{align}
The auxilliary fields are computed to be:
\begin{align}
F_U^\dagger&=-\frac{\partial W}{\partial U}
=-h_t^{(5)}\delta(y)QH_2-\partial_yU'
\\
F_{U'}^\dagger
&=-\frac{\partial W}{\partial U'}=\partial_yU\\
\delta(y)F_Q^\dagger
&=-\frac{\partial W}{\partial Q}=-h_t^{(5)}\delta(y)UH_2\\
\delta(y)F_{H_2}^\dagger
&=-\frac{\partial W}{\partial H_2}=-h_t^{(5)}\delta(y)UQ%-\mu \delta(y)H_1
\end{align}
The interaction and mass terms become 
\begin{align}
\begin{split}
\mathcal{L}_{int}=
&-\left|h_t^{(5)}\delta(y)QH_2+\partial_yU' \right|^2
-\left|\partial_yU \right|^2
-\delta(y)\left| h_t^{(5)}UH_2\right|^2
-\delta(y)\left|h_t^{(5)}UQ 
\right|^2\\
&-u\partial_yu'-h_t^{(5)}\delta(y)H_2qu-h_t^{(5)}\delta(y)Uqh_2-h_t^{(5)}
\delta(y)Qh_2u
+\text{h.c.}
\end{split}
\end{align}

Supersymmetry breaking is performed using the Scherk-Schwarz (SS)~\cite{SS}
boundary conditions based on the subgroup $U(1)_R$ which survives after the
orbifold $S^1/\mathbb{Z}_2$ action. Using the $R$-symmetry $U(1)_R$ with
parameter $\omega$ to impose different boundary conditions for bosons and
fermions inside the 5D $N=1$ multiplets, one obtains for the $n$th KK mode of
gauge bosons and chiral fermions living in the bulk the compactification mass
$n/R$, while for the $n$th KK mode of gauginos and supersymmetric partners of
chiral fermions living in the bulk, the mass $(n+\omega)/R$. Of course, the
chiral multiplets localized on the brane do not acquire any tree-level 
supersymmetry breaking mass.

After SS-breaking, the bulk hypermultiplets 
have the following Fourier expansions:
\begin{align}
u&=\frac{1}{\sqrt {\pi R}}\sum_{k=-\infty}^\infty\cos(ky/R)u^{(k)}\\
u'&=\frac{1}{\sqrt {\pi R}}\sum_{k=-\infty}^\infty\sin(ky/R)u^{(k)}\\
U&=\frac{1}{\sqrt {\pi R}}\sum_{k=-\infty}^\infty\cos((k+\omega)y/R)U^{(k)}\\
U'&=\frac{1}{\sqrt {\pi R}}\sum_{k=-\infty}^\infty\sin((k+\omega)y/R)U^{(k)}
\end{align}
where we chose to combine the two half-towers to give a single tower.

Plugging these expressions into the Lagrangian and performing the 
$y$-integration from $0$ to $\pi R$ we get
\begin{align}
\begin{split}
\label{lagrangiano}
\mathcal{L}=
&\sum_k\left\{\partial_\mu U^{(k)\dagger}\partial^\mu U^{(k)}
-\left(m_k^{B}  \right)^2U^{(k)\dagger}U^{(k)}  \right\}
+\partial_\mu Q^\dagger\partial^\mu Q
+%
\partial_\mu H_2^\dagger\partial^\mu H_2
\\
&+\sum_k \left\{i\bar u^{(k)} \bar\sigma^\mu\partial_\mu u^{(k)}
-\frac{1}{2}m_k^Fu^{(k)}u^{(k)}-\frac{1}{2}m_k^F\bar u^{(k)}\bar u^{(k)}  
\right\}
+i\bar q \bar\sigma^\mu\partial_\mu q
+i\bar h_2 \bar\sigma^\mu\partial_\mu h_2
\\
&-h_t^2\sum_k Q^\dagger Q H_2^\dagger H_2 
 - h_t^2\sum_{k,l}\left\{U^{(k)\dagger}U^{(l)} H_2^\dagger H_2
+U^{(k)\dagger}U^{(l)} Q^\dagger Q  \right\}\\
&-h_t\sum_k\left\{H_2 q u^{(k)}+U^{(k)}q h_2+Q h_2u^{(k)}+m_k^B
U^{(k)\dagger} Q H_2
+\text{h.c.}  \right\}
\end{split}
\end{align}
where we have defined $h_t^2=h^{(5)2}_t/\pi R$, $m_k^F=k/R$ and 
$m_k^B=(k+\omega)/R$ and made use of the ``representation'' 
$\delta(0)=(\pi R)^{-1}\sum_{-\infty}^{\infty} 1$. 
Had we kept two separate half towers of fermions one could 
combine the nonzero modes to give a Dirac fermion.

\section{One-loop contribution to the Higgs mass}
\label{section1loop}

In this section we compute, for the sake of completeness and to fix some
notations, the one-loop contribution to the Higgs
mass. We define the propagators of the bulk fields with end points at
$y=0$ as:
\begin{align}
D^\omega(p)&=\sum_{n=-\infty}^\infty
\frac{1}{p^2-\left(\frac{n+\omega}{R}\right)^2}
=\frac{\pi R}{2p}\bigl(\cot(\pi(pR+\omega))+\cot(\pi(pR-\omega))\bigr)\\
\widetilde D^\omega(p)&=\sum_{n=-\infty}^\infty
\frac{\frac{n+\omega}{R}}{p^2-\left(\frac{n+\omega}{R}\right)^2}
=-\frac{\pi R}{2}\bigl(\cot(\pi(pR+\omega))-\cot(\pi(pR-\omega))\bigr)
\end{align}

For Euclidean momenta ($p\rightarrow ip$)
\begin{align}\begin{split}
D^\omega_E(p) 
&=-\frac{\pi R}{2p}\bigl(\coth(\pi(pR+i\omega))+\coth(\pi(pR-i\omega))\bigr)\\
\widetilde D^\omega_E(p)
&=\frac{\pi R}{2i}\bigl(\coth(\pi(pR+i\omega))-\coth(\pi(pR-i\omega))\bigr)
\end{split}\end{align}
the propagators have the asymptotic expansions
\begin{align}
\label{expansion1}
D^\omega_E(p) 
&=-\frac{\pi R}{p}\left(1+2~e^{-2\pi Rp}\cos(2\pi\omega)\right)\\
\widetilde D^\omega_E(p)
&=2\pi R~e^{-2\pi Rp}\sin(2\pi\omega)
\label{expansion2}
\end{align}
From Eq.~(\ref{expansion1}) we can see that
$D^\omega_E(p)$ has a leading term $\sim 1/p$ which is independent of 
$\omega$ while the remaining part decays exponentially. This reflects
the soft nature of the supersymmetry breaking by the SS mechanism. 

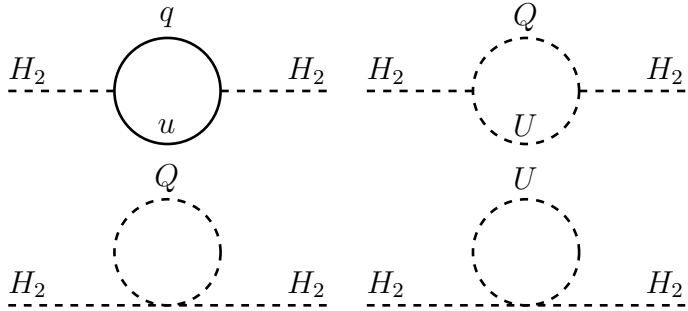
\begin{figure}[H]
\begin{center}
\begin{picture}(120,60)(0,0)
\CArc(60,20)(20,0,360) 
\DashLine(0,20)(40,20)3 
\DashLine(80,20)(120,20)3 
\Text(0,28)[l]{$H_2$}
\Text(120,28)[r]{$H_2$}
\Text(60,7)[c]{$u$}
\Text(60,48)[c]{$q$}
\end{picture}
\quad 
\begin{picture}(120,60)(0,0)
\DashCArc(60,20)(20,0,360)3
\DashLine(0,20)(40,20)3 
\DashLine(80,20)(120,20)3 
\Text(0,28)[l]{$H_2$}
\Text(120,28)[r]{$H_2$}
\Text(60,7)[c]{$U$}
\Text(60,48)[c]{$Q$}
\end{picture}

\begin{picture}(120,60)(0,0)
\DashLine(0,0)(120,0)3
\DashCArc(60,20)(20,0,360)3 
\Text(0,8)[l]{$H_2$}
\Text(120,8)[r]{$H_2$}
\Text(60,48)[c]{$Q$}
\end{picture}
\quad 
\begin{picture}(120,60)(0,0)
\DashLine(0,0)(120,0)3
\DashCArc(60,20)(20,0,360)3 
\Text(0,8)[l]{$H_2$}
\Text(120,8)[r]{$H_2$}
\Text(60,48)[c]{$U$}
\end{picture}
\end{center}
\caption{The diagrams contributing to the Higgs mass at one loop.}
\label{1loop}
\end{figure}

At one loop there are four diagrams which are shown in Fig.~\ref{1loop}.
They contribute~\cite{delgado} to the Higgs mass as
\begin{align}
\begin{split}
m_H^2&=iN_ch_t^2\intop \frac{d^4q}{(2\pi)^4} \left\{ -2D^0(q) 
+\sum_k\frac{\left(m_k^B\right)^2}{q^2\left( q^2-\left(m_k^B\right)^2 \right)}
+\sum_k\frac{1}{q^2}+D^\omega(q)\right\} \\
&=-\frac{N_ch_t^2}{4\pi^2}\intop_0^\infty dq 
\left\{ D^\omega_E(q)-D^0_E(q) \right\}
\end{split}
\end{align}

The difference $D^\omega_E(q)-D^0_E(q)$ decays exponentially 
(as can be seen from Eq.~(\ref{expansion1})) so the result is finite. 
The integral has been evaluated in Refs.~\cite{antoniadis,delgado,quiros}:
\begin{equation}
m_H^2=\frac{N_ch_t^2}{16\pi^4R^2}[
Li_3(e^{-2 i \pi \omega})+Li_3(e^{2 i \pi \omega})-2\zeta(3)]
\end{equation}
There is a cancellation of tree level infinities between the second and 
third diagrams. 
This cancellation always occurs according to the diagrammatical rule
\begin{equation}
\begin{picture}(70,25)(0,7)
\DashLine(10,0)(20,10)2
\DashLine(10,20)(20,10)2
\DashLine(20,10)(50,10)2
\DashLine(60,20)(50,10)2
\DashLine(60,0)(50,10)2
\Text(8,21)[r]{$H_2$}
\Text(8,-1)[r]{$Q$}
\Text(62,21)[l]{$H_2$}
\Text(62,-1)[l]{$Q$}
\end{picture}
\
+
\
\begin{picture}(40,25)(0,7)
\DashLine(10,20)(30,0)2
\DashLine(10,0)(30,20)2
\Text(8,21)[r]{$H_2$}
\Text(8,-1)[r]{$Q$}
\Text(32,21)[l]{$H_2$}
\Text(32,-1)[l]{$Q$}
\end{picture}
=
-ih_t^2q^2D^\omega(q)
\label{nodelta0}
\end{equation}
\vspace{0pt}

\noindent
The corresponding rule will be to drop all diagrams containing 
$\delta(0)$ and replace in the remaining ones 
\begin{equation}
\sum_{k}\frac{m_k^2}{q^2-m_k^2}\rightarrow q^2 D^\omega(q)
\end{equation}

\section{Two-loop contribution to the Higgs mass}
\label{section2loop}

In this section we compute the leading two-loop contributions to the Higgs
mass coming from the top/stop sector, as we did at one-loop in section
\ref{section1loop}, i.e. we consider $\mathcal{O}(h_t^4)$ corrections.
We use for that the Lagrangian (\ref{lagrangiano}). 
Inclusion of either gauge couplings or
a $\mu\neq 0$ term will not alter the qualitative conclusions of our analysis.
We will comment more on that in section 5.

In order to display explicitly the various cancellations
we group the genuine 
two-loop diagrams as shown in figures \ref{vanishing diagrams} 
to \ref{irreducible}, and the one-loop diagrams with counterterm insertions
in figures \ref{ctdeltaQ} and \ref{ctdeltaU}.  
In all cases external legs correspond to $H_2$ fields.

\subsection{Two-loop diagrams}
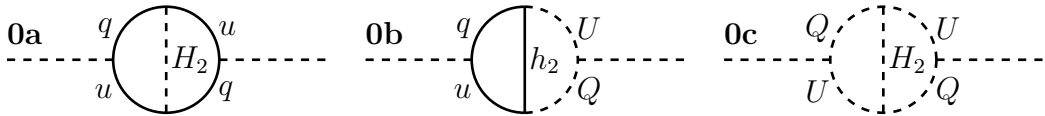
\begin{figure}[H]
\begin{center}
\begin{picture}(120,60)(0,0)
\Text(0,30)[l]{\bf 0a}
\CArc(60,20)(20,0,360)
\DashLine(60,0)(60,40)3 
\DashLine(0,20)(40,20)3 
\DashLine(80,20)(120,20)3 
\Text(40,32)[r]{$q$}
\Text(80,32)[l]{$u$}
\Text(40,8)[r]{$u$}
\Text(80,8)[l]{$q$}
\Text(62,20)[l]{$H_2$}
\end{picture}
\quad
\begin{picture}(120,60)(0,0)
\Text(0,30)[l]{\bf 0b}
\CArc(60,20)(20,90,270)
\DashCArc(60,20)(20,-90,90)3
\Line(60,0)(60,40)
\DashLine(0,20)(40,20)3 
\DashLine(80,20)(120,20)3 
\Text(40,32)[r]{$q$}
\Text(80,32)[l]{$U$}
\Text(40,8)[r]{$u$}
\Text(80,8)[l]{$Q$}
\Text(62,20)[l]{$h_2$}
\end{picture}
\quad
\begin{picture}(120,60)(0,0)
\Text(0,30)[l]{\bf 0c}
\DashCArc(60,20)(20,0,360)3
\DashLine(60,0)(60,40)3 
\DashLine(0,20)(40,20)3 
\DashLine(80,20)(120,20)3 
\Text(40,32)[r]{$Q$}
\Text(80,32)[l]{$U$}
\Text(40,8)[r]{$U$}
\Text(80,8)[l]{$Q$}
\Text(62,20)[l]{$H_2$}
\end{picture}
\end{center}
\caption{Diagrams which vanish identically.}
\label{vanishing diagrams}
\end{figure}

First there are the three diagrams
shown in figure~\ref{vanishing diagrams}.
The third one does actually not exist since there is no corresponding
contraction providing the corresponding propagators (fields are complex). 
In all diagrams containing
fermions, the terms proportional to the masses cancel, since the
masses are zero for $q$ and $h$ and for the case of the $u^{(k)}$
cancellation occurs due to symmetric summation over an odd function of
$k$. For this reason diagrams {\bf 0a} and {\bf 0b} are zero when summed over.

Contractions are done using Weyl fermions. The relevant propagators are:
\begin{equation}
\left<T(\psi_\alpha\bar\psi_\dalpha)\right>(p)=\frac{p_m\sigma^m_
{\alpha\dalpha}}{p^2-m^2}  
\end{equation}
\begin{equation}
\left<T(\psi^\alpha\bar\psi^\dalpha)\right>(p)=\frac{p^m\bar\sigma_m^
{\dalpha\alpha}}{p^2-m^2}  
\end{equation}
and the contraction rules:
\begin{align}\begin{split}
\Tr \sigma^m \bar\sigma^n &= -2 \eta^{mn}\\
\Tr \sigma^m \bar\sigma^n\sigma^r \bar\sigma^s&=-2(\eta^{mn}
\eta^{rs}-\eta^{mr}\eta^{ns}+\eta^{ms}\eta^{nr})
\end{split}\end{align}
are used.

There will be two kinds of cancellation between bosons and fermions:
\begin{itemize}
\item
If a loop involves only particles which are localized on the brane,
there will be the usual supersymmetric cancellations of quadratic
divergences. Supersymmetry breaking does not affect these particles (both
fermions and bosons remain massless).
\item
If a loop involves particles moving in the bulk, one expects cubic
divergences by power counting. In the supersymmetric theory
($\omega=0$) these terms are cancelled exactly by the sfermion. For
$\omega\neq 0$ fermions and bosons get different masses. However after
summation over the modes (integration over the loop fifth momentum) the
$\omega$-dependence lies in the propagators $D^\omega$ and $D^0$ whose
difference is exponentially suppressed for large euclidian four-momentum 
(see Eq.~(\ref{expansion1})). So the cancellation is no longer exact but 
leaves over a UV-finite piece. 
\end{itemize}

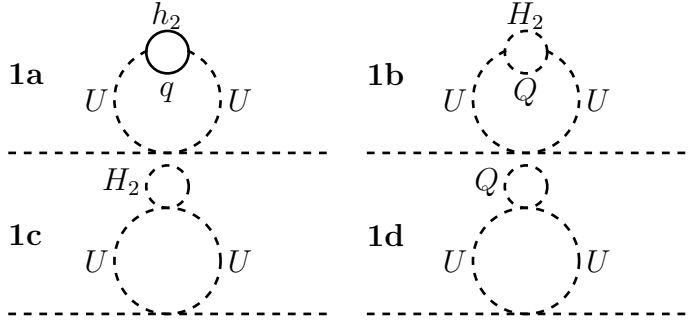
\begin{figure}[H]
\begin{center}
\begin{picture}(120,60)(0,0)
\Text(0,30)[l]{\bf 1a}
\DashLine(0,0)(120,0)3
\DashCArc(60,20)(20,112,428)3
\CArc(60,38)(8,0,360)
\Text(60,52)[c]{$h_2$}
\Text(60,23)[c]{$q$}
\Text(38,20)[r]{$U$}
\Text(83,20)[l]{$U$}
\end{picture}
\quad
\begin{picture}(120,60)(0,0)
\Text(0,30)[l]{\bf 1b}
\DashLine(0,0)(120,0)3
\DashCArc(60,20)(20,112,428)3
\DashCArc(60,38)(8,0,360)3   
\Text(60,52)[c]{$H_2$}
\Text(60,23)[c]{$Q$}
\Text(38,20)[r]{$U$}
\Text(83,20)[l]{$U$}
\end{picture}
\\
\begin{picture}(120,60)(0,0)
\Text(0,30)[l]{\bf 1c}
\DashLine(0,0)(120,0)3
\DashCArc(60,20)(20,0,360)3 
\DashCArc(60,48)(8,0,360)3
\Text(50,50)[r]{$H_2$}
\Text(38,20)[r]{$U$}
\Text(83,20)[l]{$U$}
\end{picture}
\quad
\begin{picture}(120,60)(0,0)
\Text(0,30)[l]{\bf 1d}
\DashLine(0,0)(120,0)3
\DashCArc(60,20)(20,0,360)3 
\DashCArc(60,48)(8,0,360)3
\Text(50,50)[r]{$Q$}
\Text(38,20)[r]{$U$}
\Text(83,20)[l]{$U$}
\end{picture}
\end{center}
\caption{Two-loop diagrams computed in Eqs.~(\ref{d1a})-(\ref{d1c}). 
Diagrams {\bf 1c} and {\bf 1d} have the same value.}
\label{class1}
\end{figure}

We use in the following the definitions $k=p+q,\ l=p-q$ and the identities
\begin{align}\begin{split}
2p\cdot q&=k^2-p^2-q^2\\&=-l^2+p^2+q^2\\
2p\cdot k&=-q^2+p^2+k^2\\
2q\cdot k&=-p^2+q^2+k^2
\end{split}\end{align}
to eliminate the scalar products and express all integrands in terms
of $p^2,\ q^2$ and $k^2$.  We will suppress the integrations and the
common factor of $h_t^4$.  The diagrams of the first group are shown
in figure~\ref{class1}.

The calculation of the diagrams in Fig.~\ref{class1} leads to expressions
$d_1=d_{1a}+d_{1b}+d_{1c}+d_{1d}$.
\begin{align}
\begin{split}
\label{d1a}
d_{1a}=-2iN_c\left[ D^\omega(q)  \right]^2 \frac{p\cdot k}{p^2k^2}&=
-iN_c\left[ D^\omega(q)  \right]^2 \left( -\frac{q^2}{p^2k^2} 
+\frac{1}{p^2}+\frac{1}{k^2} \right)\\
&=-iN_c\left[ D^\omega(q)  \right]^2 \left( -\frac{q^2}{p^2k^2} 
+\frac{2}{p^2} \right)
\end{split}
\end{align}
The possibility of the renaming $k\rightarrow p$ becomes obvious if one 
writes the integration over $k$ explicitly as $\intop d^4k~\delta(p+q-k)$ 
and observes that the sharp cutoff has to be introduced symmetrically for 
all three momenta. 
\begin{equation}
\label{d1b}
d_{1b}=iN_c\left[ \widetilde D^\omega(q)  \right]^2\frac{1}{k^2p^2}
\end{equation}
\begin{equation}
\label{d1c}
d_{1c}=d_{1d}=iN_c\left[ D^\omega(q)  \right]^2\frac{1}{p^2}
\end{equation}
Assembling the diagrams of Fig.~\ref{class1} we get
\begin{equation}
d_1=
iN_c\left[ D^\omega(q)  \right]^2  \frac{q^2}{p^2k^2} 
+iN_c\left[ \widetilde D^\omega(q)  \right]^2\frac{1}{k^2p^2}
\end{equation}
Note the supersymmetric cancellation of the quadratic divergences in the 
$p$-integration. The cancellation involves only loops in the brane and 
it is identical to the one that would occur in an unbroken theory.  

\begin{figure}[H]
\begin{center}
\begin{picture}(120,60)(0,0)
\Text(0,30)[l]{\bf 2a}
\DashLine(0,20)(40,20)3
\DashLine(80,20)(120,20)3
\DashCArc(60,20)(20,112,428)3
\CArc(60,38)(8,0,360)
\Text(60,52)[c]{$h_2$}
\Text(60,23)[c]{$q$}
\Text(40,32)[r]{$U$}
\Text(81,32)[l]{$U$}
\Text(60,7)[c]{$Q$}
\end{picture}
\quad
\begin{picture}(120,60)(0,0)
\Text(0,30)[l]{\bf 2b}
\DashLine(0,20)(40,20)3
\DashLine(80,20)(120,20)3
\DashCArc(60,20)(20,0,360)3 
\DashCArc(60,48)(8,0,360)3
\Text(50,50)[r]{$H_2$}
\Text(40,32)[r]{$U$}
\Text(81,32)[l]{$U$}
\Text(60,7)[c]{$Q$}
\end{picture}
\quad
\begin{picture}(120,60)(0,0)
\Text(0,30)[l]{\bf 2c}
\DashLine(0,20)(40,20)3
\DashLine(80,20)(120,20)3
\DashCArc(60,20)(20,0,360)3 
\DashCArc(60,48)(8,0,360)3
\Text(50,50)[r]{$Q$}
\Text(40,32)[r]{$U$}
\Text(81,32)[l]{$U$}
\Text(60,7)[c]{$Q$}
\end{picture}
\\
\begin{picture}(120,60)(0,0)
\Text(0,30)[l]{\bf 2d}
\DashLine(0,20)(40,20)3
\DashLine(80,20)(120,20)3
\DashCArc(60,20)(20,112,428)3
\DashCArc(60,38)(8,0,360)3
\Text(60,52)[c]{$H_2$}
\Text(60,23)[c]{$Q$}
\Text(40,32)[r]{$U$}
\Text(81,32)[l]{$U$}
\Text(60,7)[c]{$Q$}
\end{picture}
\quad
\begin{picture}(120,60)(0,0)
\Text(0,30)[l]{\bf 2e}
\DashLine(0,20)(40,20)3
\DashLine(80,20)(120,20)3
\DashCArc(60,20)(20,0,360)3 
\DashCArc(40,0)(20,0,90)3
\Text(60,48)[c]{$Q$}
\Text(40,8)[r]{$Q$}
\Text(80,8)[l]{$U$}
\Text(60,20)[c]{$H_2$}
\end{picture}
\quad
\begin{picture}(120,60)(0,0)
\Text(0,30)[l]{\bf 2f}
\DashCArc(60,20)(20,0,360)3 
\DashLine(0,20)(120,20)3 
\Text(60,28)[c]{$H_2$}
\Text(60,7)[c]{$Q$}
\Text(60,48)[c]{$Q$}
\end{picture}
\end{center}
\caption{Two-loop diagrams computed in Eqs.(\ref{d2a})-(\ref{d2d}). 
Diagrams {\bf 2b} and {\bf 2c} are equal. The last three diagrams combine 
according to the rule (\ref{nodelta0}). Note there is indeed a combinatorial 
factor of 2 for diagram {\bf 2e}, as needed.}
\label{class2}
\end{figure}
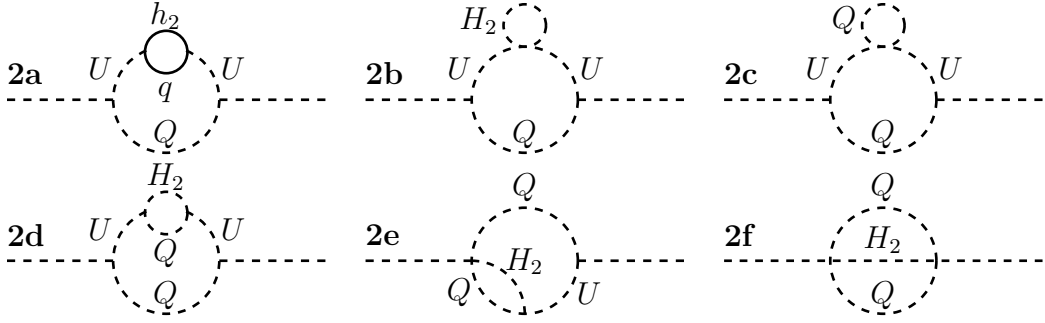

Now consider the class of diagrams of figure~\ref{class2}.

\begin{align}
\label{d2a}
d_{2a}=-2iN_c\left[ \widetilde D^\omega(q)  \right]^2\frac{p\cdot k}{q^2p^2k^2}
&=-iN_c\left[ \widetilde D^\omega(q)  \right]^2\left( \frac{1}{q^2p^2}
-\frac{1}{p^2k^2}+\frac{1}{q^2k^2}  \right)\\
&=-iN_c\left[ \widetilde D^\omega(q)  \right]^2\left( \frac{2}{q^2p^2}
-\frac{1}{p^2k^2}\right)
\end{align}

\begin{equation}
\label{d2b}
d_{2b}=d_{2c}=iN_c\left[ \widetilde D^\omega(q)  \right]^2\frac{1}{q^2p^2}
\end{equation}
\begin{equation}
\label{d2d}
d_{2d}+d_{2e}+d_{2f}=iN_c\left[ D^\omega(q)  \right]^2\frac{q^2}{p^2k^2}
\end{equation}

Adding these we get a cancellation of the quadratic divergence in the 
$p$-integration:
\begin{equation}
d_2= iN_c\left[ D^\omega(q)  \right]^2 \frac{q^2}{p^2k^2}+
iN_c\left[ \widetilde D^\omega(q)  \right]^2
\frac{1}{k^2p^2} 
\end{equation}
Again this cancellation is the unmodified supersymmetric cancellation.

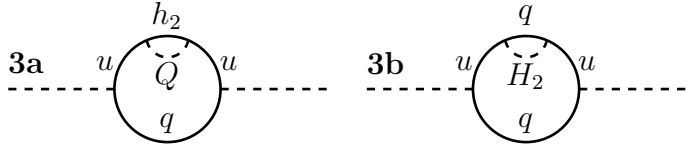
\begin{figure}[H]
\begin{center}
\begin{picture}(120,60)(0,0)
\Text(0,30)[l]{\bf 3a}
\CArc(60,20)(20,0,360) 
\DashCArc(60,40)(8,190,350)3
\DashLine(0,20)(40,20)3 
\DashLine(80,20)(120,20)3 
\Text(60,25)[c]{$Q$}
\Text(60,7)[c]{$q$}
\Text(60,48)[c]{$h_2$}
\Text(40,30)[r]{$u$}
\Text(80,30)[l]{$u$}
\end{picture}
\quad
\begin{picture}(120,60)(0,0)
\Text(0,30)[l]{\bf 3b}
\CArc(60,20)(20,0,360) 
\DashCArc(60,40)(8,190,350)3
\DashLine(0,20)(40,20)3 
\DashLine(80,20)(120,20)3 
\Text(60,25)[c]{$H_2$}
\Text(60,7)[c]{$q$}
\Text(60,48)[c]{$q$}
\Text(40,30)[r]{$u$}
\Text(80,30)[l]{$u$}
\end{picture}
\end{center}
\caption{Two-loop diagrams computed in Eq.~(\ref{d3}).}
\label{class3}
\end{figure}

The next contribution comes from correction of the $u$ propagator 
(see figure~\ref{class3}).

\begin{align}\begin{split}
\label{d3}
d_{3a}=d_{3b}&=-2iN_c\left[ D^0(q) \right]^2\frac{q\cdot p\ q^2}{p^2q^2l^2}
=-iN_c\left[D^0(q) \right]^2\left( -\frac{1}{p^2}+ \frac{1}{k^2}
+\frac{q^2}{p^2k^2}\right)\\
&=-iN_c\left[D^0(q) \right]^2
\frac{q^2}{p^2k^2}
\end{split}\end{align}

Adding $d_1$ to $d_3$ we get
\begin{equation}
d_1+d_2+d_3=2iN_c\left(\left[ D^\omega(q)  \right]^2- \left[ D^0(q)  
\right]^2\right) 
\frac{q^2}{p^2k^2}  
+2iN_c\left[ \widetilde D^\omega(q)  \right]^2
\frac{1}{k^2p^2}
\end{equation}
Now the $q$-integration is completely finite. The remaining
logarithmic divergence of the $p$-integration will be associated to
the wave-function renormalization (WFR) of the bulk-field.

\begin{figure}[H]
\begin{center}
\begin{picture}(120,60)(0,0)
\Text(0,30)[l]{\bf 4a}
\DashLine(0,0)(120,0)3
\DashCArc(60,20)(20,112,428)3
\CArc(60,38)(8,0,360)
\Text(60,52)[c]{$h_2$}
\Text(60,23)[c]{$u$}
\Text(38,20)[r]{$Q$}
\Text(83,20)[l]{$Q$}
\end{picture}
\quad
\begin{picture}(120,60)(0,0)
\Text(0,30)[l]{\bf 4b}
\DashLine(0,20)(40,20)3
\DashLine(80,20)(120,20)3
\DashCArc(60,20)(20,112,428)3
\CArc(60,38)(8,0,360)
\Text(60,52)[c]{$h_2$}
\Text(60,23)[c]{$u$}
\Text(40,32)[r]{$Q$}
\Text(81,32)[l]{$Q$}
\Text(60,7)[c]{$U$}
\end{picture}
\quad
\begin{picture}(120,60)(0,0)
\Text(0,30)[l]{\bf 4c}
\DashLine(0,0)(120,0)3
\DashCArc(60,20)(20,0,360)3 
\DashCArc(60,48)(8,0,360)3
\Text(50,50)[r]{$U$}
\Text(38,20)[r]{$Q$}
\Text(83,20)[l]{$Q$}
\end{picture}
\quad
\begin{picture}(120,60)(0,0)
\Text(0,30)[l]{\bf 4d}
\DashLine(0,20)(40,20)3
\DashLine(80,20)(120,20)3
\DashCArc(60,20)(20,0,360)3 
\DashCArc(60,48)(8,0,360)3
\Text(50,50)[r]{$U$}
\Text(40,32)[r]{$Q$}
\Text(81,32)[l]{$Q$}
\Text(60,7)[c]{$U$}
\end{picture}
\quad
\begin{picture}(120,60)(0,0)
\Text(0,30)[l]{\bf 4e}
\DashLine(0,0)(120,0)3
\DashCArc(60,20)(20,0,360)3 
\DashCArc(60,48)(8,0,360)3
\Text(50,50)[r]{$H_2$}
\Text(38,20)[r]{$Q$}
\Text(83,20)[l]{$Q$}
\end{picture}
\quad
\begin{picture}(120,60)(0,0)
\Text(0,30)[l]{\bf 4f}
\DashLine(0,0)(120,0)3
\DashCArc(60,20)(20,112,428)3
\DashCArc(60,38)(8,0,360)3
\Text(60,52)[c]{$H_2$}
\Text(60,23)[c]{$U$}
\Text(38,20)[r]{$Q$}
\Text(83,20)[l]{$Q$}
\end{picture}
\quad
\begin{picture}(120,60)(0,0)
\Text(0,30)[l]{\bf 4g}
\DashLine(0,20)(40,20)3
\DashLine(80,20)(120,20)3
\DashCArc(60,20)(20,0,360)3 
\DashCArc(60,48)(8,0,360)3
\Text(50,50)[r]{$H_2$}
\Text(40,32)[r]{$Q$}
\Text(81,32)[l]{$Q$}
\Text(60,7)[c]{$U$}
\end{picture}
\quad
\begin{picture}(120,60)(0,0)
\Text(0,30)[l]{\bf 4h}
\DashLine(0,20)(40,20)3
\DashLine(80,20)(120,20)3
\DashCArc(60,20)(20,112,428)3
\DashCArc(60,38)(8,0,360)3
\Text(60,52)[c]{$H_2$}
\Text(60,23)[c]{$U$}
\Text(40,32)[r]{$Q$}
\Text(81,32)[l]{$Q$}
\Text(60,7)[c]{$U$}
\end{picture}
\end{center}
\caption{Two-loop diagrams computed in Eqs.~(\ref{d4a})-(\ref{d4e}).
They have to be combined according to rule (\ref{nodelta0}), 
i.e. {\bf 4a+4b}, {\bf 4c+4d}, {\bf 4e+4f+4g+4h}.}
\label{class4}
\end{figure}
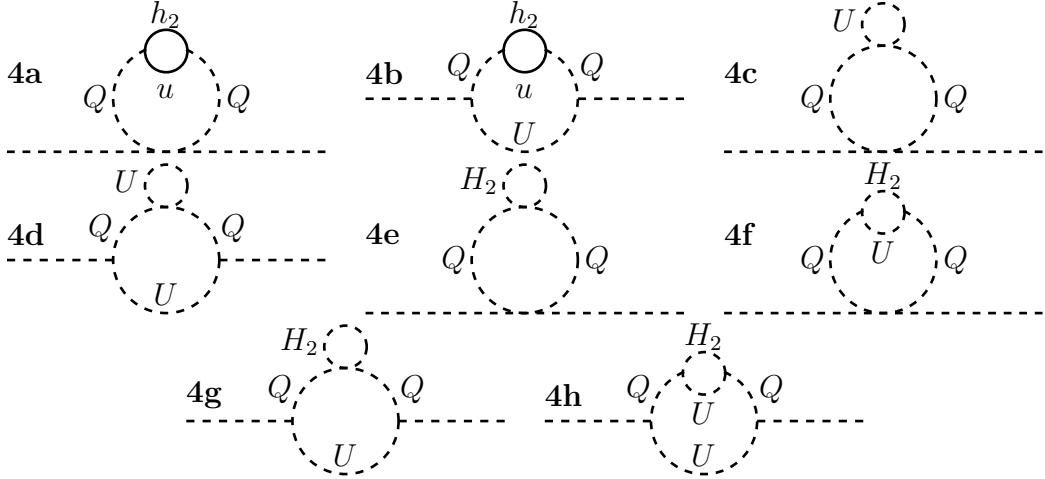

Next we consider the diagrams from figure~\ref{class4}.

All of them are infinite by themselves 
(having $\delta(0)$ vertices or divergent sums) but combining them as 
below these divergences cancel:
\begin{equation}
\label{d4a}
d_{4a}+d_{4b}=-2iN_c D^\omega(q)D^0(p)\frac{p\cdot k}{q^2k^2}
=-iN_cD^\omega(q)D^0(p)
\left( -\frac{1}{k^2}+\frac{p^2}{q^2k^2}+\frac{1}{q^2}  \right)
\end{equation}
\begin{equation}
\label{d4c}
d_{4c}+d_{4d}=iN_cD^\omega(q)D^\omega(p)\frac{1}{q^2}
\end{equation}
\begin{equation}
\label{d4e}
d_{4e}+d_{4f}+d_{4g}+d_{4h}=iN_cD^\omega(q)D^\omega(p)\frac{p^2}{q^2k^2}
\end{equation}

Summing these up we find a supersymmetric cancellation of a cubic 
divergence in the integration over $p$:
\begin{equation}
d_4=
iN_cD^\omega(q)\left( D^\omega(p)-D^0(p)  \right)\left( \frac{p^2}{q^2k^2} +\frac{1}{q^2} \right)+iN_cD^\omega(q)D^0(p)\frac{1}{k^2}
\end{equation}

\begin{figure}
\begin{center}
\begin{picture}(120,60)(0,0)
\Text(0,30)[l]{\bf 5a}
\CArc(60,20)(20,0,360) 
\DashCArc(60,40)(8,190,350)3
\DashLine(0,20)(40,20)3 
\DashLine(80,20)(120,20)3 
\Text(60,25)[c]{$U$}
\Text(60,7)[c]{$u$}
\Text(60,48)[c]{$h_2$}
\Text(40,30)[r]{$q$}
\Text(80,30)[l]{$q$}
\end{picture}
\quad
\begin{picture}(120,60)(0,0)
\Text(0,30)[l]{\bf 5b}
\CArc(60,20)(20,0,360) 
\DashCArc(60,40)(8,190,350)3
\DashLine(0,20)(40,20)3 
\DashLine(80,20)(120,20)3 
%%\Text(120,28)[r]{$H$}
\Text(60,25)[c]{$H_2$}
\Text(60,7)[c]{$u$}
\Text(60,48)[c]{$u$}
\Text(40,30)[r]{$q$}
\Text(80,30)[l]{$q$}
\end{picture}
\end{center}
\caption{Two-loop diagrams computed in Eqs.~(\ref{d5a})-(\ref{d5b}).}
\label{class5}
\end{figure}
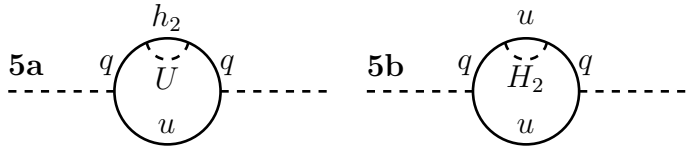

We also have the corrections to the fermionic ($q$) propagator 
(see figure~\ref{class5}):
\begin{equation}
\label{d5a}
d_{5a}=-2iN_cD^\omega(p)D^0(q)\frac{q\cdot k}{q^2k^2}=
-iN_cD^\omega(p)D^0(q)\left( -\frac{p^2}{q^2k^2}
+\frac{1}{k^2}+\frac{1}{q^2}  \right)
\end{equation}
\begin{equation}
\label{d5b}
d_{5b}=-2iN_cD^0(q)D^0(p)\frac{p\cdot q}{q^2l^2}=-iN_cD^0(q)D^0(p)
\left( -\frac{1}{q^2} + \frac{1}{k^2}+\frac{p^2}{q^2k^2} \right)
\end{equation}

\noindent
They combine to give a supersymmetric cancellation of the cubic 
divergence in the integration over $p$:
\begin{align}\begin{split}
d_5=&-iN_cD^0(q)\left( D^\omega(p)- D^0(p) \right)\left(
-\frac{p^2}{q^2k^2}+\frac{1}{q^2} \right)\\ 
&-iN_cD^0(q)\left(
D^\omega(p) +D^0(p)\right)\frac{1}{k^2}
\end{split}\end{align}

Summing up the two contributions we get:
\begin{align}\begin{split}
d_4+d_5=
&iN_c\left( D^\omega(q)-D^0(q)  \right)
\left( D^\omega(p)-D^0(p)  \right)\frac{1}{q^2}\\
+&iN_c\left( D^\omega(q)+D^0(q)  \right)
\left( D^\omega(p)-D^0(p)  \right)\frac{p^2}{q^2k^2}\\
+&iN_c\left(  
D^\omega(q)D^0(p)
-D^0(q)\left( D^\omega(p)  +D^0(p)\right) \right)\frac{1}{k^2}
\label{tempresult2}
\end{split}
\end{align}

There is a linear divergence in the $p$-integration coming from the
last term. It is associated with the WFR of the matter fields living on
the brane and will be cancelled by the usual supersymmetric
counterterm.  

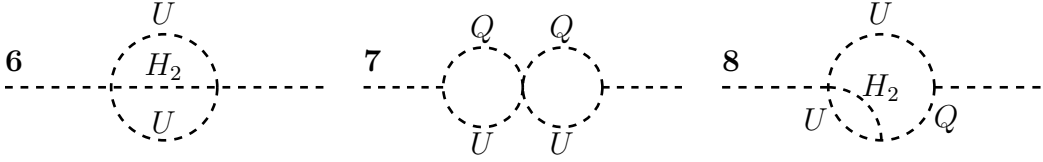
\begin{figure}[H]
\begin{center}
\begin{picture}(120,60)(0,0)
\Text(0,30)[l]{\bf 6}
\DashCArc(60,20)(20,0,360)3 
\DashLine(0,20)(120,20)3 
\Text(60,28)[c]{$H_2$}
\Text(60,7)[c]{$U$}
\Text(60,48)[c]{$U$}
\end{picture}
\quad
\begin{picture}(120,60)(0,0)
\Text(0,30)[l]{\bf 7}
\DashLine(0,20)(30,20)3
\DashLine(90,20)(120,20)3
\DashCArc(45,20)(15,0,360)3 
\DashCArc(75,20)(15,0,360)3
\Text(45,-1)[c]{$U$}
\Text(75,-1)[c]{$U$}
\Text(45,42)[c]{$Q$}
\Text(75,42)[c]{$Q$}
\end{picture}
\quad
\begin{picture}(120,60)(0,0)
\Text(0,30)[l]{\bf 8}
\DashLine(0,20)(40,20)3
\DashLine(80,20)(120,20)3
\DashCArc(60,20)(20,0,360)3 
\DashCArc(40,0)(20,0,90)3
\Text(60,48)[c]{$U$}
\Text(40,8)[r]{$U$}
\Text(80,8)[l]{$Q$}
\Text(60,20)[c]{$H_2$}
\end{picture}
\end{center}
\caption{Last three two-loop contributions, computed in 
Eqs.~(\ref{d6})-(\ref{d8}). Diagrams {\bf 7} and {\bf 8} are UV-finite}
\label{irreducible}
\end{figure}

To make the $q$-integration finite we look at the three
remaining diagrams (see figure~\ref{irreducible})

\begin{align}
\label{d6}
d_6&=iN_cD^\omega(q)D^\omega(p)\frac{1}{k^2}\\
\label{d7}
d_7&=iN_c^2\widetilde D^\omega(p) \widetilde D^\omega(q)\frac{1}{q^2p^2}\\
\label{d8}
d_8&=2iN_c\widetilde D^\omega(p) \widetilde D^\omega(q)\frac{1}{k^2p^2}
\end{align}
$d_7$ and $d_8$ are UV finite according to the expansion (\ref{expansion2}), 
while $d_6$ is exactly the missing piece in equation (\ref{tempresult2}).

We end up with:
\begin{align}\begin{split}
-N_ch_t^4\intop \frac{d^4p}{(2\pi)^4}~\frac{d^4q}{(2\pi)^4}\ \biggl\{
2&\left(\left[ D^\omega_E(q)  \right]^2- \left[ D^0_E(q)  \right]^2  \right) 
\frac{q^2}{p^2k^2}-2\left[ \widetilde D^\omega_E(q)  \right]^2\frac{1}{k^2p^2}
\biggr.\\
+&\left( D^\omega_E(q)-D^0_E(q)  \right)\left( D^\omega_E(p)-D^0_E(p)  \right)
\frac{1}{q^2}\\
+&\left( D^\omega_E(q)+D^0_E(q)  \right)\left( D^\omega_E(p)-D^0_E(p)  \right)
\frac{p^2}{q^2k^2}\\
+&\left( D^\omega_E(q)
-D^0_E(q)\right)\left( D^\omega_E(p)  +D^0_E(p)\right) \frac{1}{k^2}\\\biggl.
-&N_c\widetilde D^\omega_E(p) \widetilde D^\omega_E(q)\frac{1}{q^2p^2}-2
\widetilde D^\omega_E(p) \widetilde D^\omega_E(q)\frac{1}{k^2p^2}\biggr\}
\label{result1}
\end{split}\end{align}

One sees that the only remaining divergences are in the
$p$-integration, more specifically there is the logarithmic divergence
in the first line and the linear divergence in the fourth. The
expression vanishes for $\omega=0$.

\subsection{The counterterms}

As in any (exact or softly broken) supersymmetric theory the 
superpotential couplings run due to the wave-function renormalization of
the superfields. 
We would like to rescale the fields which couple to the brane in a 
manifestly supersymmetric way. To this end we rescale the fields as follows: 
\begin{alignat}{3}
Q&\rightarrow\left(1-\frac{\delta_Q}{2}\right)Q 
&q&\rightarrow\left(1-\frac{\delta_Q}{2}\right)q\\
H_2&\rightarrow\left(1-\frac{\delta_{H_2}}{2}\right)H_2
&h_2&\rightarrow\left(1-\frac{\delta_{H_2}}{2}\right)h_2\\
U&\rightarrow\left(1-\frac{\delta_U}{2}\pi R~\delta(y)\right)U
&u&\rightarrow\left(1-\frac{\delta_U}{2}\pi R~\delta(y)\right)u\\
\partial_yU'&\rightarrow\left(1+\frac{\delta_U}{2}\pi R~\delta(y)\right)
\partial_yU'
\quad&\partial_yu'&\rightarrow\left(1+\frac{\delta_U}{2}\pi R~\delta(y)\right)
\partial_yu'
\end{alignat}
$\partial_yU$ and $U'$ do not couple to the brane and thus will not
receive a brane rescaling. 
We define also a rescaled (renormalized) coupling as 
\begin{equation}
h_t=\left(1- \frac{\delta_Q}{2}- \frac{\delta_{H_2}}{2}- 
\frac{\delta_U}{2}\pi R~\delta (0)  \right)h_t^{bare}
\end{equation}

By inspection of equation (\ref{Superfields}) it is obvious that this 
leaves the Lagrangian unchanged if expressed in rescaled quantities. 
The different sign for $U$ and $\partial_y U'$ guarantees that the first 
term of the superpotential remains unaltered. 

The following counterterms now appear in the 4D Lagrangian:
\begin{align}
\begin{split}
{\cal L}_{CT}=&-\delta_Q\partial_\mu Q^\dagger\partial_\mu Q
-\delta_{H_2}\partial_\mu H_2^\dagger\partial_\mu H_2
-\delta_U\pi R\left( \partial_\mu U^\dagger\partial_\mu U 
+ \partial_y U^{\prime\dagger}\partial_y U'\right)_{y=0}\\
&-\delta_Qq\sigma^\mu\partial_\mu\bar q
-\delta_{H_2}h_2\sigma^\mu\partial_\mu\bar h_2
-\delta_U\pi R\left(u\sigma^\mu\partial_\mu\bar u\right)_{y=0}
\\
&-\delta_Qh_t^2\pi R\left (U^\dagger U\right)_{y=0}H_2^\dagger H_2
-\delta_{H_2}h_t^2\pi R\left (U^\dagger U\right)_{y=0}Q^\dagger Q\\
&-\delta_Uh_t^2(\pi R\delta(0))^2Q^\dagger QH_2^\dagger H_2
+\delta_Uh_t\pi R\delta(0)\left( QH_2\left(\partial_y U' \right)_{y=0} 
+ \text{\text{h.c.}} \right)\\
=&-\delta_Q\partial_\mu Q^\dagger\partial_\mu Q
-\delta_{H_2}\partial_\mu H_2^\dagger\partial_\mu H_2
-\delta_U\sum_{k,\ l}\partial_\mu U^{(k)\dagger}\partial_\mu U^{(l)}
-\delta_U\sum_{k,\ l}m_k U^{(k)\dagger}m_l U^{(l)}\\
&-\delta_Qq\sigma^\mu\partial_\mu\bar q
-\delta_{H_2}h_2\sigma^\mu\partial_\mu\bar h_2
-\delta_U\sum_{k,\ l}u^{(k)}\sigma^\mu\partial_\mu\bar u^{(l)}
\\
&-\delta_Qh_t^2\sum_{k,\ l}U^{(k)\dagger} U^{(l)}H_2^\dagger H_2
-\delta_{H_2}h_t^2\sum_{k,\ l}U^{(k)\dagger} U^{(l)}Q^\dagger Q\\
&-\delta_Uh_t^2(\pi R\delta(0))^2Q^\dagger QH_2^\dagger H_2
+\delta_Uh_t\pi R\delta(0)\left( \sum_k m_k U^{(k)}QH_2 + \text{h.c.} \right)
\end{split}
\end{align}

The values of $\delta_U$ and $\delta_Q$ are determined by the infinite
parts of the one-loop two point functions at nonvanishing momentum $q$.
One finds:
\begin{align}\begin{split}
\left<U_k^\dagger U^{\phantom\dagger}_l\right>&=h_t^2\intop
\frac{d^4p}{(2\pi)^4}\frac{m^B_km^B_l+q^2}{p^2k^2}\\
&=ih_t^2\left( m^B_km^B_l+q^2  \right) \frac{1}{8\pi^2}\log \left( \frac{\Lambda^2}{\mu^2} \right)+\text{finite}
\end{split}\end{align}
\begin{align}\begin{split}
\left<Q^\dagger Q\right>&=h_t^2\intop\frac{d^4p}{(2\pi)^4}\left[ D^\omega(p)-D^0(p)  \right]\left( 1+\frac{p^2}{k^2} \right)+\frac{q^2}{k^2}D^0(p)\\
&=ih_t^2q^2 \frac{R\Lambda}{8\pi} +\text{finite}\label{QQ}
\end{split}\end{align}
where $\mu$ is an infrared cutoff. 
The infinities can be absorbed if one chooses
\begin{align}
\delta_U&=h_t^2\frac{1}{8\pi^2}\label{deltaU}\log \left( \frac{\Lambda^2}{\mu^2}\right)\\
\delta_Q&=h_t^2\frac{R\Lambda}{8\pi}\label{deltaQ}
\end{align}

\begin{figure}[H]
\begin{center}
\begin{picture}(120,60)(0,0)
\CArc(60,20)(20,0,360) 
\DashLine(0,20)(40,20)3 
\DashLine(80,20)(120,20)3 
\Text(60,7)[c]{$q$}
\Text(40,30)[r]{$u$}
\Text(80,30)[l]{$u$}
\ctu(60,40)
\end{picture}
\quad
\begin{picture}(120,60)(0,0)
\DashCArc(60,20)(20,0,360)3
\DashLine(0,20)(40,20)3 
\DashLine(80,20)(120,20)3 
\Text(60,7)[c]{$Q$}
\Text(40,30)[r]{$U$}
\Text(80,30)[l]{$U$}
\ctu(60,40)
\end{picture}
\quad
\begin{picture}(120,60)(0,0)
\DashCArc(60,20)(20,0,360)3
\DashLine(0,20)(40,20)3 
\DashLine(80,20)(120,20)3 
\Text(60,7)[c]{$Q$}
\Text(60,48)[c]{$U$}
\ctu(40,20)
\end{picture}
\\
\begin{picture}(120,60)(0,0)
\DashLine(0,0)(120,0)3
\DashCArc(60,20)(20,0,360)3 
\Text(38,20)[r]{$U$}
\Text(83,20)[l]{$U$}
\ctu(60,40)
\end{picture}
\quad
\begin{picture}(120,60)(0,0)
\DashLine(0,0)(120,0)3
\DashCArc(60,20)(20,0,360)3 
\Text(60,48)[c]{$Q$}
\ctu(60,0)
\end{picture}
\end{center}
\caption{Counterterm diagrams, computed in Eq.~(\ref{comp10}), 
proportional to $\delta_U$.}
\label{ctdeltaU}
\end{figure}
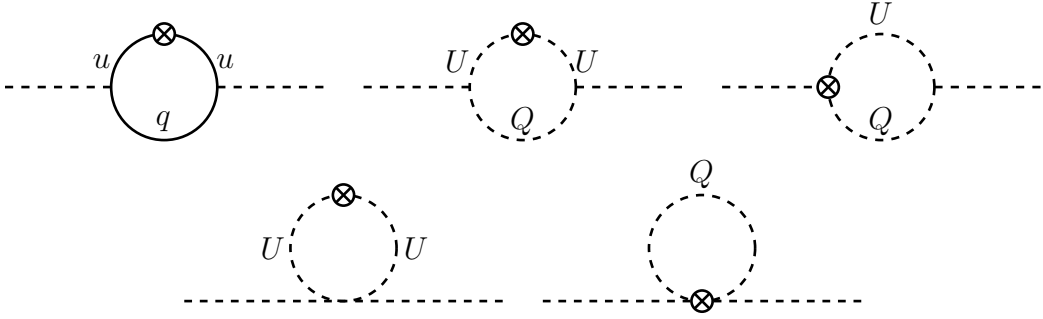

There are now the additional diagrams shown in figures \ref{ctdeltaQ} 
and \ref{ctdeltaU} ($\delta_U$ counterterms are denoted by 
$\ctu(3,3)\ \ $ and $\delta_Q$ counterterms by $\ctq(3,3)\ \ $).

The diagrams in figure~\ref{ctdeltaQ} contribute:
\begin{equation}
\label{comp9}
2\delta_Q \left( D^\omega(q)- D^0(q) \right)
\end{equation}
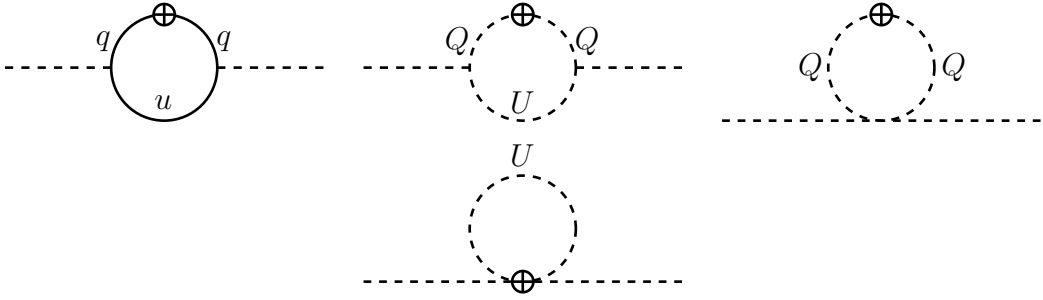
\begin{figure}[H]
\begin{center}
\begin{picture}(120,60)(0,0)
\CArc(60,20)(20,0,360) 
\DashLine(0,20)(40,20)3 
\DashLine(80,20)(120,20)3 
\Text(60,7)[c]{$u$}
\Text(40,30)[r]{$q$}
\Text(80,30)[l]{$q$}
\ctq(60,40)
\end{picture}
\quad
\begin{picture}(120,60)(0,0)
\DashCArc(60,20)(20,0,360)3
\DashLine(0,20)(40,20)3 
\DashLine(80,20)(120,20)3 
\Text(60,7)[c]{$U$}
\Text(40,30)[r]{$Q$}
\Text(80,30)[l]{$Q$}
\ctq(60,40)
\end{picture}
\quad
\begin{picture}(120,60)(0,0)
\DashLine(0,0)(120,0)3
\DashCArc(60,20)(20,0,360)3 
\Text(38,20)[r]{$Q$}
\Text(83,20)[l]{$Q$}
\ctq(60,40)
\end{picture}
\\
\begin{picture}(120,60)(0,0)
\DashLine(0,0)(120,0)3
\DashCArc(60,20)(20,0,360)3 
\Text(60,48)[c]{$U$}
\ctq(60,0)
\end{picture}
\end{center}
\caption{Counterterm diagrams, computed in Eq.~(\ref{comp9}), 
proportional to $\delta_Q$.}
\label{ctdeltaQ}
\end{figure}

The diagrams of figure \ref{ctdeltaU} give 
(observe the factor of two for the third diagram):
\begin{multline}
\label{comp10}
\delta_U\sum_{k,\ l}\left(
\frac{1}{q^2}\frac{m_k}{q^2-m_k^2}\frac{m_l}{q^2-m_k^2}(q^2+m_km_l)+
2\frac{1}{q^2}\frac{m_k^2}{q^2-m_k^2}
\right.\\
\left.
+\frac{1}{q^2-m_k^2}\frac{1}{q^2-m_k^2}(q^2+m_km_l)+
\frac{1}{q^2}
\right)-2\delta_Uq^2\left[  D^0(q) \right]^2
\\
=\delta_U\sum_{k,\ l}\frac{1}{q^2}\frac{1}{q^2-m_k^2}\frac{1}{q^2-m_k^2}
\biggl(
 (q^2+m_km_l)m_km_l 
+2(q^2-m_l^2)m_k^2
\biggr.
\\
\biggl.
+q^2(q^2+m_km_l) +(q^2-m_k^2)(q^2-m_l^2)
\biggr)-2\delta_Uq^2\left[  D^0(q) \right]^2\\
=2\delta_U\left( q^2\left[D^\omega(q)\right]^2 -q^2\left[D^0(q)\right]^2+\left[  \widetilde D^\omega(q)  \right]^2\right)
\end{multline}

So the total counterterm contribution is:
\begin{equation}
-N_ch_t^2\intop \frac{d^4q}{(2\pi)^4}~2\delta_Q \left( D^\omega_E(q)- 
D^0_E(q) \right)-2\delta_U\left( q^2\left[D^\omega_E(q)\right]^2 
-q^2\left[D^0_E(q)\right]^2-\left[  \widetilde D^\omega_E(q)  \right]^2\right)
\end{equation}

Adding them to the divergent terms of (\ref{result1}) we get
\begin{align}\begin{split}
-N_ch_t^2
\intop \frac{d^4q}{(2\pi)^4}\ \biggl\{2&\left(
q^2\left[ D^\omega_E(q)  \right]^2- q^2
\left[ D^0_E(q)  \right]^2  -\left[ \widetilde D^\omega_E(q)  
\right]^2\right)\left(h_t^2\intop \frac{d^4p}{(2\pi)^4}\ 
\frac{1}{k^2p^2}-\delta_U\right)\biggr.\\
+&\left( D^\omega_E(q)
-D^0_E(q)\right)\left(h_t^2\intop \frac{d^4p}{(2\pi)^4}\ 
\left( D^\omega_E(p)  +D^0_E(p)\right) \frac{1}{k^2}+2\delta_Q\right)
\end{split}\end{align}
which is finite if the counterterms are chosen according to 
(\ref{deltaU}, \ref{deltaQ}). 

\section{Conclusions}
\label{conclusion}

In this paper we have made an explicit two-loop calculation of the Higgs
mass in the five dimensional model of Ref.~\cite{quiros}, where the fifth
dimension is compactified on the orbifold $S^1/\mathbb{Z}_2$ and supersymmetry
is broken in the bulk by means of Scherk-Schwarz boundary conditions. 
We have considered only (top) Yukawa couplings in the calculation ($h_t$) and
disregarded gauge couplings ($g$). We have also considered $\mathcal{O}(\mu^0)$
results, which amounts to disregard the $\mu$-parameter in the superpotential.
Our result is that no Higgs mass counterterm is generated at two-loop and the
only divergent terms in the two-loop correction can be absorbed by the 
renormalization group running of $h_t$ between the cutoff $\Lambda$ and 
the scale where the theory becomes four dimensional, $1/R$. The latter can be
interpreted as threshold effects at $1/R$ when matching the 4D theory below
it with the complete KK theory above $1/R$. As a result our explicit
calculation shows that the Higgs mass is finite and calculable to the
considered order in perturbation theory.

Inclusion of gauge couplings has been done at one-loop, $\mathcal{O}(g^2)$
terms, for an arbitrary gauge group~\cite{antoniadis,delgado}. At two-loop we
would have $\mathcal{O}(g^2 h_t^2)$ and $\mathcal{O}(g^4)$ terms. We expect
divergent terms which are canceled by supersymmetry or by counterterms from
supersymmetric wave-function renormalization, which can be absorbed by the
running of gauge and Yukawa couplings, yielding again a finite result.

Inclusion of the $\mu$-term (a supersymmetric mass) in the
superpotential would introduce divergences, even at the one-loop
level, that can be interpreted as the supersymmetric running of the
tree level Higgs mass terms, while the analysis becomes much more
involved since the Higgs mass $\mathcal{M}^2$ becomes a two-by-two
matrix. All infinities, in this case, are expected to be absorbed by
supersymmetry as well as by the running of gauge and Yukawa couplings,
and of the $\mu$-parameter.  The resulting absence of Higgs mass
counterterms should hold in this case.  At one-loop, we have checked
that the field rescaling $\delta_{H_2}=N_c \delta_Q$ is indeed
sufficient to render $\mathcal{M}^2$ finite.

\section*{Acknowledgements} 
The work of AD was supported 
by the Spanish Education Office (MEC) under an \emph{FPI} scholarship.   
The work of GG was supported by the DAAD.

\end{document}